\documentclass[aps,prd,superscriptaddress,nofootinbib]{revtex4}

\usepackage{amsmath,amssymb}
\usepackage[bookmarks=false,colorlinks=true]{hyperref}
\usepackage{color}
\usepackage{tikz}
\usepackage{latexsym}
\def\DR{\rm I\kern-1.45pt\rm R}
\def\DC{\kern2pt {\hbox{\sqi I}}\kern-4.2pt\rm C}

\newcommand{\ba}{\begin{array}}
\newcommand{\ea}{\end{array}}
\newcommand{\be}{\begin{equation}}
\newcommand{\ee}{\end{equation}}
\newcommand{\bea}{\begin{eqnarray}}
\newcommand{\eea}{\end{eqnarray}}

\begin{document}
\title{$\mathbb{CP}^N$-Rosochatius  system, superintegrability,  supersymmetry}

\author{Evgeny Ivanov}
\email{eivanov@theor.jinr.ru}
\affiliation{Bogoliubov Laboratory of Theoretical Physics, Joint Institute for Nuclear Research, 141980, Dubna, Moscow Reg., Russia}
\author{Armen Nersessian}
\email{arnerses@yerphi.am}
\affiliation{Yerevan Physics Institute, 2 Alikhanian Brothers St., Yerevan  0036 Armenia}
%\affiliation{Yerevan State University, 1 Alex Manoogian St., Yerevan, 0025, Armenia}
%\affiliation{Bogoliubov Laboratory of Theoretical Physics, Joint Institute for Nuclear Research, 141980, Dubna, Moscow Reg., Russia}
\author{Hovhannes Shmavonyan}
\email{hovhannes.shmavonyan@mail.yerphi.am}
\affiliation{Yerevan Physics Institute, 2 Alikhanian Brothers St., Yerevan  0036 Armenia}

\begin{abstract}
We  propose new superintegrable mechanical system on the
complex projective space $\mathbb{CP}^N$ involving a potential term  together with coupling to a constant magnetic fields. This system can be viewed
as a  $\mathbb{CP}^N$-analog of both the flat singular oscillator
and its spherical analog known as ``Rosochatius system''.
  We find its constants of motion and calculate their  (highly nonlinear)  algebra. We also present its classical and quantum solutions.
The system belongs to the class of  ``K\"ahler oscillators" admitting  $SU(2|1)$ supersymmetric extension.
We  show that, in the absence of magnetic field and with the special choice of the characteristic parameters, one can construct
$\mathcal{N}=4, d=1$  Poincar\'e supersymmetric extension of the system considered.
\end{abstract}

\maketitle
\section{Introduction}
The ($D$-dimensional) isotropic oscillator and the relevant Coulomb problem play a pivotal role among other textbook examples of $D$-dimensional integrable systems.
  They are distinguished by the ``maximal superintegrability" property, which is the existence of $2D-1$ functionally independent constants of motion \cite{perelomov}.
The rational Calogero model with oscillator potential \cite{Calogero},  being a nontrivial generalization  of isotropic oscillator,  is also  maximally superintegrable \cite{woj}.
Moreover, Calogero model with Coulomb potential is superintegrable too \cite{rapid}.
All these systems, being originally defined on a plane, admit the maximally superintegrable deformations to the spheres (see Ref. \cite{higgs} for the spherical generalizations
 of the oscillator and Coulomb problem, and Ref. \cite{rapid} for the  Calogero-oscillator and Calogero-Coulomb ones).
The integrable spherical generalizations  of anisotropic oscillator \cite{Vahagn}, Stark-Coulomb and two-center Coulomb problems \cite{2center}
are  also known.

In contrast to the spherical extensions, the generalizations to other curved spaces have not  attracted much attention so far.
The only exception  seems to be  the isotropic oscillator on  the complex/quaternionic spaces considered in Ref. \cite{CPosc,QPN}.
These systems reveal an important feature: they remain superintegrable after coupling to a constant magnetic/BPST instanton field,
though cease to be maximally superintegrable. One may pose a question:\\

{\sl How to construct the {\it superintegrable} generalizations of Calogero-oscillator and Calogero-Coulomb models on complex and quaternionic projective spaces?}\\

In this paper we make first steps toward the answer. Due to the complexity of the problem
 we restrict our attention to the simplest particular case. Namely,
 we construct  the superintegrable $\mathbb{CP}^N$-generalization of
 the $N$-dimensional singular oscillator (the simplest rational Calogero-oscillator model) which is defined by the Hamiltonian
\be\label{SWR}
H_{SW}=\sum_{a=1}^{N}\Big(\frac{p_a^2}{2}+\frac{g_a^2}{2x_a^2}+\frac{\omega^2x_a^2}{2}\Big),\qquad
\{p_a,x_b\}=\delta_{ab},\quad \{p_a,p_b\}=\{x_a,x_b\}=0. \ee This
model is less trivial than it looks at first sight: it has  a
variety of hidden constants of motion  which form a nonlinear
symmetry algebra and endow the system with the maximal
superintegrability property. Its extensive studies were initiated
more than fifty years ago by Smorodinsky
 with collaborators \cite{SW} and are  continuing up to now (see, e.g., \cite{fazlul} and references therein).
 Sometimes this model is referred to as Smorodinsky-Winternitz system, though it was known  for many years.

The maximally superintegrable spherical counterpart  of the Smorodinsky-Winternitz system is defined by the Hamiltonian  suggested  by Rosochatius in 1877 \cite{Ros}

\be
H_{Ros}=\frac{1}{2}\sum_{a,b=1}^N(\delta_{ab}-\frac{x_ax_b}{r^2_0})p_ap_b+ \sum_{a=1}^N(\frac{\omega^2_ar^2_0}{x^2_a}+\frac{\omega^2r^2_0 x^2_a}{2x^2_0})%-\frac{\omega^2r^2_0}{2},
,\qquad x^2_a+x^2_0=r^2_0\,.
\label{Rosin}\ee
It is a particular case of the integrable systems  obtained by restricting the free particle and oscillator  systems to a sphere.
It was studied  by many authors from different viewpoints, including its re-invention  as a superintegrable spherical generalization of Smorodinsky-Winternitz system \cite{mack,pogosyanSph,gns}.
Rosochatius  model, as well as its hybrid with the Neumann model suggested in 1859 \cite{neumann}, attract a stable interest for years due to their relevance to
a wide circle of physical and mathematical problems. Recently, the Rosochatius-Neumann system was encountered, while studying strings \cite{arut},
extreme black hole geodesics \cite{gns,sheikh} and  Klein-Gordon equation in curved backgrounds \cite{evnin}.\\

In this paper we  propose a superintegrable generalization of Rosochatius (and Smorodinsky-Winternitz) system on the complex projective space $\mathbb{CP}^N$.
It is  defined by  the Hamiltonian \footnote{ Hereafter we use the notation $z\bar z\equiv \sum_{c=1}^N z^c{\bar z}^c$}
 \be\label{Hamr0}
{\cal H}_{Ros}=\frac1{r^2_0}{(1+z\bar z)(\delta^{a\bar b}+z^a{\bar z}^b)\pi_a\bar\pi_b } + r^2_0(1+z\bar z)(\omega^2_0+ \sum_{a=1}^N\frac{\omega^2_a}{z^a{\bar z}^a})-r^2_0\sum_{i=0}^N\omega^2_i,
%\qquad\{\pi_a, z^b\}=\{\bar\pi_a, \bar z^b\}=\delta_{\bar a}^{\bar b},
\ee
and  by the Poisson brackets providing the interaction with a constant magnetic field of the magnitude $B$
 \be \label{com}
 \{\pi_a, z^b\}=\delta_a^b,\quad \{\bar\pi_a, \bar z^b\}=\delta_{\bar a}^{\bar b},\quad \{\pi_a,\bar\pi_b\}=\imath B{r^2_0}\left(\frac{\delta_{a\bar b}}{1+z\bar z}-\frac{\bar z^a z^b}{(1+z\bar z)^2}\right).
 \ee
We will call it {\bf \sl $\mathbb{CP}^N$-Rosochatius system}\footnote{Despite the fact that $U(N)$ symmetry is explicitly broken in (\ref{Hamr0}) (down to $U(1)^N$), hereafter
we use the $U(N)$ covariant notation, such that $\pi_a$ and $\pi_{\bar b}$ transform, respectively, as $\bar z^a$ and $z^b$ and there are three equivalent forms
of the $U(N)$ invariant tensor, $\delta_{a\bar b}, \delta_a^b$ and $\delta_{\bar b}^{\bar a}$.}.

Reducing  this $2N$-dimensional system by the action of  $N$ manifest $U(1)$ symmetries, $z^a\to{\rm e}^{\imath\kappa_a}z^a,\pi_a\to{\rm e}^{-\imath\kappa_a}\pi_a $,
we recover the $N$-dimensional Rosochatius system \eqref{Rosin} (see {\sl Section 3}).

 On the other hand, rescaling the coordinates and momenta as $r_0z^a\to z^a, \pi_a/r_0\to\pi_a$ and taking the limit $r_0\to \infty, \omega_a\to 0$ with
 $r^2_0\omega_a=g_a$ kept finite, we arrive  at the   so-called ``$\mathbb{C}^N$-Smorodinsky-Winternitz system" \cite{shmavon}
\be
\mathcal{H}_{SW}=\sum_{a=1}^N\left(\pi_a\bar\pi_a +\omega^2_0 z^a\bar z^a+\frac{g^2_a}{z^a\bar z^a}\right),\qquad
 \{\pi_a, z^b\}=\delta_a^b,\quad \{\bar\pi_a, \bar z^b\}=\delta_{\bar a}^{\bar b},\quad \{\pi_a,\bar\pi_b\}=\imath B\delta_{a\bar b}\,.
%\Omega= dz\wedge d\pi +d{\bar z}\wedge d{\bar\pi} + \imath Bdz\wedge d{\bar z}.
\label{SW0}\ee
Since the reductions of $\mathbb{CP}^N$-Rosochatius system yield superintegrable systems, it is quite natural that it proves to be superintegrable on its own.

We will show that  $\mathbb{CP}^N$-Rosochatius system belongs to the class of  ``K\"ahler oscillators" \cite{CPosc,Kahlerosc}  which
admit $SU(2|1)$ supersymmetrization (or a `weak $\mathcal{N}=4$" supersymmetrization, in terminology of Smilga \cite{smilga}).
A few years ago it was found that these systems naturally arise within  the appropriate $SU(2|1), d=1$ superspace formalism
developed in a series of papers \cite{sidorov}. This research was partly motivated by
the study of  the field theories with curved rigid analogs of Poincar\'e supersymmetry \cite{csa}.
In the absence of the background magnetic field  and for the special choice of the parameters $\omega_i$, the  $\mathbb{CP}^N$-Rosochatius system
admits $\mathcal{N}=4, d=1$ Poincar\'e  supersymmetric extension.

Finally, note that  $\mathbb{C}^N$-Smorodinsky-Winternitz system \eqref{SW0} can be interpreted as
a set of $N$ two-dimensional ring-shaped oscillators interacting with a constant magnetic field orthogonal to the plane.
%In the case  when all  $g_a$ are equal, it describes ensemble of $N$ particle  in quantum ring.
As opposed to  \eqref{SW0},  the $\mathbb{CP}^N$-Rosochatius system does not split into a set of $N$ two-dimensional decoupled systems.
Instead, it can be interpreted as describing  {\sl interacting} particles with a position-dependent mass in the two-dimensional quantum rings
(along the lines of ref. \cite{chp,chp1,chp2}).

To summarize, the {\sl $\mathbb{CP}^N$-Rosochatius system} suggested is of interest from many points of view.
Its study is the subject of the remainder of this paper. It is organized as follows.\\

In {\sl Section 2} we review the main properties of the complex projective space $\mathbb{CP}^N$,
the simplest  related systems like $\mathbb{CP}^N$-Landau problem and the $\mathbb{CP}^N$-oscillator,
and then derive the potential specifying the $\mathbb{CP^N}$-Rosochatius system.

In  {\sl Section 3} we present  classical $\mathbb{CP}^N$-Rosochatius model in a constant magnetic field and find
that, in addition to  $N$ manifest $U(1)$ symmetries,  this system possesses additional $2N-1$ functionally-independent second-order
constants of motion. The latter property  implies the (non-maximal) superintegrability of the model considered.
We present the explicit expressions of  the constants of motion and calculate their algebra. We also show that
the reduction of  $\mathbb{CP}^N$-Rosochatius model by manifest $U(1)$ symmetries reproduces the original $N$-dimensional ($\mathbb{S}^N$-) Rosochatius system.

In   {\sl Section 4} we   separate the variables and find  classical solutions of $\mathbb{CP}^N$-Rosochatius model.

In  {\sl Section 5} we  study quantum $\mathbb{CP}^N$-Rosochatius system and find its spectrum which depends on $N+1$ quantum numbers,
as well as the relevant  wavefunctions.

In  {\sl Section 6} we construct $\mathcal{N}=4$ supersymmetric extensions of $\mathbb{CP}^N$-Rosochatius system.

In  {\sl Section 7} we give an account of  open problems and possible  generalizations.\\

In the subsequent consideration we  put,  for simplicity,  $r_0=1$.

\section{Preliminaries: complex projective spaces}
In this Section we  present the basic properties of  complex projective space $\mathbb{CP}^N$,
briefly describe the Landau problem and the oscillator  on this  space, and construct
$\mathbb{CP}^N$-analog of Rosochatius system.

The $N$-dimensional complex projective space is a space of complex rays in the $(N+1)$-dimensional complex Euclidian space
$(\mathbb{C}^{N+1},\; \sum_{i=0}^N du^i d\bar u^i)$, with $u^i$ being homogeneous coordinates of the complex projective space.
 Equivalently, it can be defined as the quotient $\mathbb{S}^{2N+1}/U(1)$, where  $\mathbb{S}^{2N+1}$ is the $(2N+1)$-dimensional sphere  embedded
 in $\mathbb{C}^{N+1}$ by the constraint $\sum_{i=1}^N u^i\bar u^i =1$.
One  can solve the latter by introducing locally ``inhomogeneous" coordinates $z^a_{(i)}$
\be
z^a_{(i)}=\frac{u^a}{u^i},\qquad{\rm with}\quad  a\neq i, u^i\neq 0.
\ee
Hence,  the full complex projective space can be covered by  $N+1$ charts marked by the indices $i=0,\ldots, N$, with the following
transition functions on the intersection  of $i$-th and $j$-th charts:
\be
z^a_{(i)}=\frac{z^a_{(j)}}{z^i_{(j)}}.%,\qquad a=0,\ldots, {\hat i},\ldots N
\label{transition}\ee
Let us endow $\mathbb{C}^{N+1}$ with the canonical Poisson brackets $\{u^i,{\bar u}^j\}=\imath\delta^{i\bar j}$, and
define, with respect to them, the $u(N+1)$ algebra formed by the generators
\be
h_{i\bar j}= {\bar u}^i u^j\,.
\ee
Reducing the manifold $\mathbb{C}^{N+1}$ by the action of the $U(1)$ group with the generator  $h_0= \sum_{i=0}^N u^i{\bar u}^i$,
 we  arrive at the $SU(N+1)$-invariant Ka\"hler structure  defined by the Fubini-Study metrics
 \begin{equation}
 \sum_{a, b=1}^N g_{a\bar b}dz^ad{\bar z}^b =\sum_{a, b=1}^N\frac{\partial^2\log(1+z\bar z)}{\partial z^a\partial\bar z^b}dz^ad{\bar z}^b =
 \sum_{a, b=1}^N \left(\frac{\delta_{ a \bar b}}{1+z \bar z}-\frac{\bar z^az^b}{(1+z \bar z)^2 }\right)dz^ad\bar z^b.
 \label{FS} \end{equation}
This metrics is obviously invariant under the passing from one chart to another. Hence, we can omit the  indices marking charts and assume,
without loss of generality,  that we are dealing with  $0$-th chart, so that the indices $a,b,c$ run from $1$ to $N$.

Being K\"ahler manifold, the complex projective space is  equipped with the Poisson brackets $\{{\bar z}^a, z^b\}_0=-\imath g^{\bar ab}$,
where $g^{\bar a b}=(1+z\bar z)(\delta^{\bar a b}+\bar z^a z^b)$ is the inverse Fubini-Study metrics.
 The $su(N+1)$ isometry of $\mathbb{CP}^N$ is generated by the  holomorphic Hamiltonian vector fields defined as the
  following momentum maps  (Killing potentials)
\be
h_{a\bar b}=\frac{\bar{z}^a z^b}{1+z\bar z},\qquad h_a=\frac{2{\bar z}^a}{1+z\bar z}.
\label{suKilling}\ee

Now, let us introduce, on the cotangent bundle of $\mathbb{C}^{N+1}$, the canonical Poisson brackets $\{p_i, u^j\}=\delta_{ij}$,
and  define the  $su(N+1)$ algebra with the generators
\be
L_{i\bar j}=\imath ( p_i  u^j-\bar p_j \bar u^i)-\frac{\delta_{i\bar j}}{N}L_0 ,\qquad {\rm where}\quad L_0=\imath \sum_{i=0}^N( p_i  u^i-\bar p_i\bar u^i).
\label{Lij}\ee
 Reducing this phase space  by the action of  generators $L_0, h_0=\sum_i u^i\bar u^i$, and finally fixing their values as $L_0=2B, h_0=1$, we arrive
 at the Poisson brackets \eqref{com} (with $r_0=1$). They describe an electrically charged particle on $\mathbb{CP}^N$
 interacting with  a constant magnetic field of the magnitude $B$ and set the corresponding  twisted symplectic structure
\be
\Omega_0=\sum_{a=1}^N
(dz^a\wedge d\pi_a +d{\bar z}^{a}\wedge d{\bar\pi}_a) + B\sum_{a,b=1}^N\imath g_{a\bar b}dz^a\wedge
d{\bar z}^b,
\label{hn}\ee
with $g_{a\bar b}$ being defined in \eqref{FS}.

 The inhomogeneous coordinates and momenta  $z^a, \pi_a$ are related to the homogeneous ones $p_i, u^i$ as  \cite{lnp}
\be
z^a=\frac{u^a}{u^0},\quad
\pi_a=\sum_{b=1}^Ng_{a\bar b}\Big(\frac{p_b}{\bar u^0}-\bar{z}^b\frac{ p_0}{\bar u^0}\Big).
\ee
The  $su(N+1)$ generators \eqref{Lij} are reduced to the following ones
\bea\label{suN1}
&J_{a\bar b}=\imath(z^b\pi_a-\bar\pi_b\bar z^a)- B\frac{\bar z^a z^b}{1+z\bar z},\quad
J_{a}=\pi_a+\bar z^a(\bar z\bar\pi)+\imath B\frac{{\bar z}^a}{1+z\bar z}%, \quad
:&
\label{su}\\
&\{J_{{\bar a} b}, J_{\bar c d}\}=
i\delta_{\bar a d}J_{\bar b c}
-i\delta_{\bar c b}J_{\bar a d},\quad \{J_{a}, {\bar J}_{b}\}=-i (J_{a\bar b}+J_0 \delta_{a\bar b}),\quad \{J_a, { J}_{b\bar c}\}=i J_b\delta_{a\bar c},&
\eea
where $J_0\equiv \sum_{a=1}^N J_{a\bar a}+B$.

With these expressions at hand we can now consider some superintegrable systems on $\mathbb{CP}^N$.
\\

{\bf $\mathbb{CP}^N$-Landau problem.}
 The $\mathbb{CP}^N$-Landau problem is defined by the symplectic structure \eqref{hn}  and the free-particle Hamiltonian identified with a  Casimir  of $su(N+1)$ algebra
%is defined by the expression
\be
\mathcal{H}_{0}=\sum_{a,b=1}^N{(1+z\bar z)(\delta^{a\bar b}+z^a{\bar z}^b)\pi_a\bar\pi_b }=\frac12\sum_{i,j=0}^{N}L_{i\bar j}L_{j\bar i}-\frac{B^2}{2}=\sum_{a=1}^NJ_a\bar J_a+\frac{\sum_{a,b=1}^NJ_{a\bar b}J_{b \bar a}+J_0^2-B^2}{2}:\quad \{\mathcal{H}_0, L_{ij}\}=0.
\ee
Its   quantization  was done, e.g., in  \cite{nair}.\\

{\bf $\mathbb{CP}^N$-oscillator.}
The $\mathbb{CP}^N$-oscillator is defined by the symplectic structure \eqref{hn} and the Hamiltonian \cite{CPosc}

\be
\mathcal{H}_{osc}= \sum_{a,b=1}^N{(1+z\bar z)(\delta^{a\bar b}+z^a{\bar z}^b)\pi_a\bar\pi_b }+
\omega^2\sum_{a=1}^N z^a\bar z^a\,.
\label{cpnosc}\ee
It respects manifest $U(N)$ symmetry with the generators $J_{a\bar b}$  \eqref{suN1}, and additional hidden symmetries  given by the proper analog of ``Fradkin tensor",
\be\label{fradkinosc}
I_{a\bar b}={J_a {\bar J}_b} +\omega^2 {\bar z}^a z^b\,.
\ee
The full  symmetry algebra of this system  reads
\bea
&
\{J_{{\bar a} b}, J_{\bar c d}\}=
\imath\delta_{\bar a d}J_{\bar b c}
-\imath\delta_{\bar c b}J_{\bar a d},\quad
\{I_{a\bar b}, J_{c\bar d}\}=\imath\delta_{a\bar d}I_{c\bar b}-
\imath\delta_{c\bar b}I_{a\bar d}& \\
 &\{I_{a \bar b}, I_{c\bar d}\}=\imath\omega^2
 \delta_{a\bar d}J_{c\bar b}-
\imath\omega^2 \delta_{c\bar b} J_{a\bar d}
-\imath I_{c\bar b}(J_{a\bar d}+J_0\delta_{a\bar d})
+\imath I_{a\bar d}(J_{c\bar b}+J_0\delta_{c\bar b})\,,&
\label{constraint}
\eea
where
$
J_0=\imath\sum_{a=1}^N(z^a\pi_a-\bar\pi_a\bar z^a)+B\frac{1}{1+z\bar z}
$.

The Hamiltonian \eqref{cpnosc} is expressed via the symmetry generators as follows
\be
{\cal H}_{osc}=\sum_{a=1}^NI_{a\bar a}+\frac 12\sum_{a,b=1}^N J_{a\bar b}J_{b \bar a}+ \frac{J_0^2-B^2}{2}.
\label{cpnoscconstants}\ee
The quantum mechanics associated with this Hamiltonian was considered in \cite{qcpn}.
In the flat limit, the $\mathbb{CP}^N$-oscillator  goes over to the $\mathbb{C}^N$-oscillator interacting with a constant magnetic field.\\

{\bf $\mathbb{CP}^N$-Rosochatius system.}
The $\mathbb{CP}^N$-oscillator, being   superintegrable system (for $N>1$), has an obvious drawback: it lacks covariance  under  transition from one chart to another.
This non-covariance becomes manifest after expressing  the Hamiltonian \eqref{cpnosc}  via the $SU(N+1)$ symmetry generators and the homogeneous coordinates $u^i$,
\be
{\cal H}_{osc}=\frac{\sum_{i,j=0}^{N} L_{i\bar j}L_{j\bar i}-B^2}{2}+\frac{\omega^2}{u^0{\bar u}^0}-\omega^2.
\label{cpnosc1}\ee
This expression  allows one to immediately construct  $(N+1)$-parameter deformation of the $\mathbb{CP}^N$-oscillator, such that it is manifestly form-invariant
under  passing from one chart to another accompanied by the appropriate change of the parameters $\omega_i$. The relevant potential is
 \be
V_{Ros}=\sum_{i=0}^N\left(\frac{\omega^2_i}{u^i{\bar u}^i}-\omega^2_i\right),\qquad {\rm with}\quad \sum_{i=0}^N u^i{\bar u}^i=1.
\label{ros}\ee
In the case when all parameters $\omega_i$ are equal, the system is  globally  defined on the complex projective space with the punctured points $u^i=0\,$.

The system with the potential \eqref{ros}  is just  the $\mathbb{CP}^N$-Rosochatius system mentioned in Introduction.
Now we turn to its investigation as the main subject of the present paper.

\section{$\mathbb{CP}^N$-Rosochatius system}
We consider   the  $N$-parameter deformation of the  $\mathbb{CP}^N$- oscillator by the potential \eqref{ros}, in what follows  referred to as the
``$\mathbb{CP}^N$-Rosochatius system". It is defined by the Hamiltonian \eqref{Hamr0} and  Poisson brackets  \eqref{com} with $r_0=1$.
Equivalently, this system can be defined by the symplectic structure \eqref{hn} and the Hamiltonian

\be\label{Ham}
{\cal H}_{Ros}=\sum_{a,b=1}^N {g^{\bar a b}\bar\pi_a\pi_b} + (1+z\bar z)\left(\omega_0^2+ \sum_{a=1}^N\frac{\omega_a^2}{z^a{\bar z}^a}\right)-\sum_{i=0}^N\omega^2_i,
\ee
where $g^{\bar a b}=(1+z\bar z)(\delta^{\bar a b}+  \bar z^a z^b)$ is the inverse Fubini-Study metrics.

The model has $N$ manifest (kinematical) $U(1)$   symmetries  with the generators
\be
J_{a\bar a}=\imath\pi_a z^a-\imath\bar{\pi}_a\bar{z}^a-B\frac{ z^a{\bar z}^a}{1+z{\bar z}}\; :\quad\{J_{a\bar a}, \mathcal{H}\}=0,
\label{u1n}\ee
and hidden symmetries with the  second-order generators $I_{ij}=(I_{0a},I_{ab})$ defined as
\be
I_{0a}=J_{0a}{\bar J}_{0\bar a} +\omega_0^2 z^a{\bar z}^a +\frac{\omega_a^2}{{\bar z}^a z^a} ,\qquad I_{ab}=J_{a\bar b}J_{b\bar a}+\omega_a^2\frac{z^b{\bar z}^b}{z^a{\bar z}^a} +
\omega_b^2\frac{z^a{\bar z}^a}{z^b{\bar z}^b}\; :  \quad\{I_{i\bar j}, \mathcal{H}\}=0\,.
\label{fradkinRos}\ee
In the homogeneous coordinates, the hidden symmetry  generators can be cast in a more succinct form
\be
I_{ij}=J_{i\bar j}J_{j\bar i}+ \omega_i^2\frac{u^j\bar u^j}{u^i\bar u^i}+\omega_j^2 \frac{u^i\bar u^i}{u^j\bar u^j} .
\ee
The relevant symmetry algebra is given by the brackets
\be
 \{J_{a\bar a}, I_{ij}\}=0, \qquad \{I_{ij}, I_{kl}\}=\delta_{jk}T_{ijl}+\delta_{ik}T_{jkl}-\delta_{jl}T_{ikl}-\delta_{il}T_{ijk}\,,
 \ee
 with
 \be
( T_{ijk})^2=2(I_{ij}-J_{i\bar i}J_{j \bar j})(I_{jk}-J_{j\bar j}J_{k \bar k})(I_{ik}-J_{i\bar i}J_{k\bar k})
 +2I_{ij}I_{ik}I_{jk}+J_{i\bar i}^2J_{j \bar j}^2J_{k\bar k}^2-(I_{jk}^2J_{i \bar i}^2+I_{ij}^2J_{k \bar k}^2+I_{ik}^2J_{j\bar j}^2)
\nonumber\ee
\be
-4(\omega_k^2I_{ij}(I_{ij} - J_{i\bar i}J_{j\bar j}) + \omega_i^2I_{jk}(I_{jk} - J_{j \bar j}J_{k \bar k}) +
    \omega_j^2I_{ik}(I_{ik} - J_{i\bar i}J_{k \bar k})) + 4\omega_j^2 \omega_k^2J_{i\bar i}^2 +
 4\omega_i^2 \omega_k^2J_{j \bar j}^2 + 4\omega_i^2 \omega_j^2J_{k\bar k}^2 + 16\omega_i^2 \omega_j^2 \omega_k^2.
 \ee
The  Hamiltonian is expressed via these generators as follows
\be
\mathcal{H}=\frac12\sum_{i=1}^{N+1}I_{ij}+\sum_{a=1}^N \omega_a^2+\frac{J_0^2-B^2}{2}=\sum_{a=1}^N I_{0a}+\sum_{a,b=1}^N\frac{I_{ab}}{2}+\sum_{a=1}^N \omega_a^2+\frac{J_0^2-B^2}{2}.
\ee

This consideration actually proves the superintegrability of the $\mathbb{CP}^N$-Rosochatius system. The number of the functionally independent
constants of motion will be counted in the end of this Section.

For sure, the   symmetry algebra written above can be found by a direct calculation of the Poisson brackets between the symmetry generators.
However, there is a more elegant and simple way to construct it. Namely, one has to consider the symmetry algebra of $\mathbb{C}^{N+1}$-Smorodinsky-Winternitz system \cite{shmavon}
with {\sl vanishing} magnetic field, and  to reduce it, by action of the generators $\sum_{i=0}^N\imath(p_iu^i-\bar p_i\bar u^i)$, $\sum_{i=0}^Nu^i\bar u^i\,$(see the previous Section),
to the symmetry algebra of $\mathbb{CP}^N$-Rosochatius system.

\subsection*{Reduction to (spherical) Rosochatius system}

In order to understand the relationship with the standard Rosochatius system (defined on the sphere) let us pass to the real canonical variables $y_a,\varphi^a$, $p_a, p_{\varphi_a}$
\be
z^a={ y_a{\rm e}^{\imath\varphi_a}},\quad \pi_a=\frac{1}{{2}}\left(p_a-{\imath}\Big(\frac{p_{\varphi_{a}}}{y_a}+\frac{B y_a}{1+y^2}\Big) \right){\rm e}^{-\imath\varphi_a} \; :\qquad\Omega=dp_a\wedge dy_a
+ dp_{\varphi_{a}}\wedge d\varphi_a\,.
%+Bd\frac{y^2_a d\varphi_a}{1+y^2}.
\label{realcoor}\ee
In these variables the Hamiltonian \eqref{Ham} is rewritten as
\be
\mathcal{H}_{Ros}=\frac{1}{4}(1+\sum_{c=1}^Ny^2_c)\left[ \sum_{a,b=1}^N( \delta_{ab} +y_ay_b) p_ap_b+ 4{\widetilde\omega}^2_0+ 4\sum_{a=1}^N\frac{{\widetilde\omega}^2_a}{y^2_a} \right]-E_0\,,
\label{realHam}\ee
where
\be
{\widetilde\omega}^2_a=\omega^2_a+\frac14 {p^2_{\varphi_a}},\quad {\widetilde\omega}^2_0=\omega^2_0+\frac14\left(B+\sum_{a=1}^Np_{\varphi_a}\right)^2,\qquad
E_0=\frac{B^2}{4}+\sum_{i=0}^N \omega^2_i.
\label{tilde}\ee
Then, performing the reduction by cyclic variables $\varphi^a$ ({\it i.e.}, by  fixing the momenta $p_\varphi^a$), we arrive at the Rosochatius system
on the sphere with $y_a=x_a/x_0$, where
  $(x_0, x_a)$ are ambient Cartesian coordinates, $\sum_{i=0}^Nx^2_i=1$:
  \bea
  x_a=\frac{y_a}{\sqrt{1+\sum_{c=1}^Ny^2_c}},\quad x_0=\frac{1}{\sqrt{1+\sum_{c=1}^Ny^2_c}}\,.
  \eea
  As was already noticed, the $\mathbb{S}^N$-Rosochatius system   is maximally superintegrable, {\it i.e.} it has $2N-1$ functionally independent constants of motion.
  From the above reduction we conclude that the $\mathbb{CP}^N$-Rosochatius system has
  $2N-1+ N=3N-1$ functionally independent integrals. Hence, it lacks $N$ integrals needed for the maximal superintegrability.

\section{Classical Solutions}
To obtain the  classical  solutions of $\mathbb{CP}^N$-Rosochatius system we   introduce the spherical coordinates through the recursion
\be
y_N=r\cos\theta_{N-1},\quad y_{\alpha}=r\sin\theta_{N-1}u_\alpha, \qquad {\rm with} \quad r=\tan\theta_{N},\qquad \sum_{\alpha=1}^{N-1}u^2_\alpha=1,
\label{sphcoor}\ee
where $y_a$ were defined by \eqref{realcoor}.
In terms of these coordinates the Hamiltonian  \eqref{realHam} takes the form
\be\label{sphRos}
\mathcal{H}_{Ros}\equiv \mathcal{I}_{N}-E_0=\frac14(1+r^2)\left((1+r^2)p^2_r +\frac{4\mathcal{I}_{N-1}(\theta)}{r^2} +4\widetilde\omega^2_0\right)-E_0,
\quad \mathcal{I}_{a}=\frac{p^2_{\theta_{a}}}{4}+\frac{\mathcal{I}_{a-1}}{\sin^2\theta_{a}} +\frac{\widetilde\omega^2_{a+1}}{\cos^2\theta_{a}},
\ee
with $E_0$, $\omega_N\equiv {\tilde\omega}_0$ defined in \eqref{tilde}, $a=1, \ldots,N$ and $\mathcal{I}_0=0$.

 Thus we singled out the complete set  of Liouville integrals $(\mathcal{H}_{Ros}, \mathcal{I}_{\alpha}, p_{\varphi_{a}})$, and separated the variables.
 It is by no means the unique choice of Liouville integrals and of the coordinate frame in which  the Hamiltonian admits the separation of variables.
 However, for our purposes it is enough to deal with any particular choice.

 With the above expressions at hand,
 we  can derive classical solutions of the system by solving  the  Hamilton-Jacobi equation
 \be
 \mathcal{H}(p_\mu=\frac{\partial S}{\partial x^\mu} , x^\mu)=E,\qquad {\rm with}\quad  x^\mu=(\theta_a,\varphi_a ),
 \quad p_\mu=(p_a,p_{\varphi_{a}}).
 \ee
 To this end, we introduce the generating function  of the form
 \be
 S_{tot}=2\sum_{a=1}^{N}S_{a}(\theta_a)+\sum_{a=1}^N p_{\varphi_a}\varphi_a\,.
 \ee
 Substituting this ansatz in the Hamilton-Jacobi equation, we immediately separate the variables and arrive at  the set of ordinary differential equations:
\be
\left(\frac{dS_{{a}}}{d\theta_{a}}\right)^2+\frac{c_{a-1}}{\sin^2\theta_{a}} +\frac{\widetilde\omega^2_{a+1}}{\cos^2\theta_{a}}
=c_a,\qquad a = 1, \ldots,N, {\quad c_N := E+E_0, \quad \widetilde\omega^2_{N+1} := \widetilde\omega^2_{0}}\,.
\ee
Solving these equations, we obtain
\be
S_{{a}}=\int d\theta_{a}\sqrt{c_{a}-\frac{c_{a-1}}{\sin^2\theta_{a}} -\frac{\widetilde\omega^2_{a+1}}{\cos^2\theta_{a}}}\,.
\ee
Thus we  have found the general solution of the Hamilton-Jacobi equation ({\it i.e.}, the solution depending on $2N$ integration constants
$c_a, \, p_{\varphi_{a}}$).

In order to get the solutions of the classical equations of motion,  we should differentiate the generating functions with respect to these integration constants
and then equate the resulting functions to some constants $t_0,\kappa_\alpha,$ and $\varphi^a_0 $,
\be
\frac{\partial S_{tot}}{\partial E}=t-t_0,\qquad \frac{\partial S_{tot}}{\partial c_\alpha}=2\sum_{b=1}^N \frac{\partial S_b}{\partial c_\alpha}=\kappa_\alpha,\quad \alpha=1,\ldots, N-1, \qquad
\frac{\partial S_{tot}}{\partial p_{\varphi_{a}}}=\varphi^a+\sum_{b=1}^{N}2\frac{\partial S_b}{\partial p_{\varphi_a}}=\varphi^a_0\,. \label{EqDif}
\ee
Introducing
\be
\xi_a :=\sin^2\theta_a,\quad \mathcal{A}_a :=\frac{c_{a}+c_{a-1}-{\widetilde\omega}^2_{a+1} }{2c_a}\,,
\ee
we obtain from \eqref{EqDif}
\be
\xi_{N}-\mathcal{A}_N=\sqrt{\mathcal{A}_{N}^2-\frac{c_{N-1}}{c_{N}}}\sin 2\sqrt{c_N}(t-t_0),\ee
  \be
  \xi_{\alpha}=\sqrt{{\cal A}_{\alpha}^2-\frac{c_{\alpha-1}}{c_{\alpha}}}\Bigg(\frac{\sin \kappa_{\alpha}(\xi_{\alpha+1}{\cal A}_{\alpha+1}
  -\frac{c_{\alpha}}{c_{\alpha+1}})+\cos \kappa_{\alpha}\sqrt{-\xi_{\alpha+1}^2
  +2\xi_{\alpha+1}{\cal A}_{\alpha+1}-\frac{c_{\alpha}}{c_{\alpha+1}}}}{\xi_{\alpha+1}\sqrt{\frac{c_{\alpha+1}}{c_{\alpha}}{\cal A}_{\alpha+1}^2-1}}\Bigg)
  +{\cal A}_{\alpha},
\ee
\be
\varphi^a-\varphi^a_0=-\frac{ p_{\varphi_a}}{4\widetilde\omega_{a+1}}\arctan \frac{2 \widetilde\omega_{a+1} \sqrt{c_{a-1} \left(\xi_{a }-1\right)-\xi_{a } \left(c_a \left(\xi _{a
   }-1\right)+\widetilde\omega_{a+1}^2\right)}}{-c_{a-1} \left(\xi_{a
   }-1\right)+c_a \left(\xi_{a }-1\right)\widetilde\omega_{a+1}^2\left(\xi_{a }+1\right)}\,.
\ee
Thereby we have derived the  explicit classical solutions  of our $\mathbb{CP}^N$-Rosochatius system.

\section{Quantization}
In order to quantize the
$\mathbb{CP}^N$-Rosochatius system
we  replace the Poisson brackets \eqref{com} by the commutators (with $r_0=1$)
\be
 [\widehat{\pi}_a, z^b]=-\imath\hbar \delta^b_a,\qquad [\widehat{\pi}_a, \widehat{\bar\pi}_b]=\hbar B\left(\frac{\delta_{a\bar b}}{1+z\bar z}-\frac{\bar z^a z^b}{(1+z\bar z)^2}\right).%g_{a\bar b}.
\ee
The appropriate quantum realization of the momenta operators reads
\be
\widehat{\pi}_a=-\imath\left(\hbar\frac{\partial }{\partial z^a} +\frac{B}{2}\frac{\bar z^a}{1+z\bar z}\right),\quad \widehat{\bar\pi}_a=-\imath\left(\hbar\frac{\partial }{\partial\bar z^a} -\frac{B}{2}\frac{\bar z^a}{1+z\bar z}\right).
\ee
Then we define the quantum Hamiltonian
\be\label{qH}
\widehat{\mathcal{H}}_{Ros}=\frac{1}{2} g^{a\bar b}\left( \widehat{\pi}_a\widehat{\bar\pi}_b+\widehat{\bar\pi}_b\widehat{\pi}_a\right) +\hbar^2(1+z\bar z)\left(\omega_0^2+ \sum_{a=1}^N\frac{\omega_a^2}{z^a{\bar z}^a}\right)-\hbar^2\sum_{i=0}^N\omega_i^2.
\ee
The kinetic term in this Hamiltonian  is written as the Laplacian on K\"ahler manifold (coupled to a magnetic field)
defined with respect to the volume element $dv_{\mathbb{CP}^N}=(1+z\bar z)^{-(1+N)}[dzd\bar z]$, while  in  the potential term we
have made the replacement $\omega_i\to\hbar \omega_i\,$.

In terms of the real coordinates $z^a=y_a{\rm e}^{\imath\varphi_a}$ this Hamiltonian reads (cf. \eqref{realHam})
\be\label{rqH}
\widehat{\mathcal{H}}_{Ros}=(1+\sum_{c=1}^Ny^2_c)\Bigg[-\frac{\hbar^2}{4}\Big(\sum_{a,b=1}^{N}(\delta_{ab}+y_a y_b)
\frac{\partial^2}{\partial y_a\partial y_b}+\sum_{a=1}^N\Big(y_a+\frac{1}{y^a}\Big)\partial_{y_a}\Big)+\widehat{\mathbb{\tilde{\omega}}}_{N+1}^2
+ \sum_{a=1}^N\frac{ \widehat{\mathbb{\tilde{\omega}}}_\alpha^2}{4y_a^2}\Bigg]-\tilde{E_0}\,.
\ee
Here  we introduced the operators
\be
\widehat{\mathbb{\tilde{\omega}}}_{N+1}^2=\Big(\frac{B}{\hbar}+\frac{1}{\hbar}\sum_{a=1}^N\widehat{p}_{\varphi_a}\Big)^2+4\omega_0^2,\qquad
\widehat{\mathbb{\tilde{\omega}}}_a^2=4{\omega}_a^2+\frac{\widehat{p}^2_{\varphi_a}}{ \hbar^2 }
\ee
with
\be
\widehat{p}_{\varphi_a}=\widehat{J}_{a\bar a}=-\imath\hbar\frac{\partial}{\partial\varphi^a}
\qquad
 \tilde{E_0}=\frac{B^2}{
4}+\hbar^2\sum_{i=0}^N\omega_i^2.
\ee
Clearly, these operators are quantum analogs of the classical quantities \eqref{tilde}.
In  the spherical coordinates \eqref{sphcoor} the Hamiltonian \eqref{rqH} takes the form
\be
\widehat{\mathcal{H}}_{Ros}=\widehat{\mathcal{I}}_N - \tilde{E}_{0},\qquad
 \widehat{\mathcal{I}}_a=-\frac{\hbar^2}{4}\Bigg((\sin\theta_{a})^{1-a}\frac{\partial}{\partial \theta_{a}}\Big((\sin \theta_a)^{a-1}\frac{\partial}{\partial \theta_{a}}\Big)
 +(a \cot \theta_{a}-\tan \theta_{a})\frac{\partial}{\partial \theta_{a}}\Bigg)+\frac{ \widehat{\mathcal{I}}_{a-1}}{\sin^2 \theta_a}
 +\frac{\hbar^2 \widehat{\mathbb{\tilde{\omega}}}_{a+1}^2}{4\cos^2\theta_{a}},
\label{spher}
\ee
where $a =1,...,N$ and  $\widehat{\mathcal{I}}_0=0$.

This prompts  us to consider the spectral problem \footnote{ In the classical limit, $\hbar\to 0,\; m_a, l_a\to\infty $, the eigenvalues $\hbar m_a$
yield $p_{\varphi_a}$ and $\hbar l_a$ yield $\sqrt{c_a}$.}
\be
 \widehat{J}_{a\bar a}\Psi=\hbar m_a\Psi, \qquad \widehat{\mathcal{I}}_a\Psi=\frac{\hbar^2}{4}l_a(l_a+2a)\Psi,
\label{sp}\ee
where $l_a$ are the appropriate ``spin'' quantum numbers, and separate the variables  by the choice of  the wavefunction in such a way that
it resolves  first $N$ equations in the above problem,
\be
\Psi=\frac{1}{({2\pi})^{N/2}}\prod_{a=1}^N\psi_{a}(\theta_{a})e^{\imath m_a\varphi_a},\quad m_a=0,\pm 1, \pm 2, \ldots
\ee
Then, passing to the variables  $\xi_{a}=\sin^2 \theta_{a}$,  we transform the reduced  spectral problem to the system of $N$ ordinary differential equations
\be
-\xi_a(1-\xi_a)\psi_{a}''+\big((a+1)\xi-a\big)\psi_{a}'+\frac{1}{4}\Bigg(\frac{l_{a-1}(l_{a-1}+2a-2)}{\xi_a}+\frac{\tilde\omega_{a+1}^2}{1-\xi_a}-l_{a}(l_{a}+2a)\Bigg)\psi_a=0.
\ee
These equations can be cast in the form of a  hypergeometric equation through the following substitution
\be
\psi(\xi_a)=\xi_a^{\frac{l_{a-1}}{2}}\Big(1-\xi_a\Big)^{\frac{\omega_{a+1}}{2}}f(\xi_a):
\ee
\be
\xi_a(1-\xi_a)f''+\Big(l_{a-1}+a-\xi\Big(l_{a-1}+a+\tilde{\omega}_{a+1}+1\Big)\Big)f'-\frac{1}{4}\Big(l_{a-1}+\tilde{\omega}_{a+1}-l_a)(l_{a-1}+\tilde{\omega}_{a+1}+l_a+2a) \Big)f=0.
\ee
The regular solution of this equation is the hypergeometric function \cite{flu}
\be
f_a(\xi)=C_0F(-n_a; l_{a-1}+\tilde{\omega}_{a+1}+a+n_a; l_{a-1}+a; \xi_a),\quad l_{a}=2n_a+l_{a-1}+\tilde{\omega}_{a+1} ,  \label{HyperG}
\ee
with
\be
 n_a=0,1,2... \;\qquad \tilde{\omega}_{a}=\sqrt{4\omega_a^2+m_a^2}.
\ee
Therefore,
$
l_N=\sum_{a=1}^N\left( 2 n_a + \tilde{\omega}_{a}\right)$,  so that
 the energy spectrum is given  by the expressions
\be
E_{n,\{m_a\}}=\frac{\hbar^2}{4}\Bigg(2n+N+\sqrt{({B}/{\hbar}+\sum_{a=1}^N m_a)^2+4\omega_0^2}+\sum_{a=1}^{N}\sqrt{4\omega_a^2+m_a^2}\Bigg)^2- \frac{B^2+\hbar^2N^2}{
4}-\hbar^2\sum_{i=0}^N\omega_i^2,
\ee
where $n=\sum_{a=1}^N n_a=0,1,\ldots$ \footnote{For the integer parameters $n_a$
the hypergeometric function \eqref{HyperG} is reduced to Jacobi polynomials. We thank
Referee for this remark.}.

Thus the spectrum of quantum $\mathbb{CP}^N$-Rosochatius system depends on $N+1$ quantum numbers. This is  in full agreement with the fact that
this system has $3N-1$  functionally independent constants of motion
(let us  remind that the spectrum of $D$-dimensional quantum mechanics with $D+K$ independent integrals of motion depends on $D-K$ quantum numbers.
E.g, the spectrum of maximally superintegrable system depends on the single (principal) quantum number).

Let us also write down the explicit expressions for the non-normalized wavefunctions and the  $\mathbb{CP}^N$ volume element
\be
\Psi_{\{n_a\},\{m_a\}}=\frac{C_0}{({2\pi})^{N/2}}\prod_{a=1}^N\xi_a^{\frac{l_{a-1}}{2}}\Big(1-\xi_a\Big)^{\frac{\omega_{a+1}}{2}}e^{\imath m_a\varphi_a} F(-n_a; l_{a-1},+\tilde{\omega}_{a+1}+a+n_a; l_{a-1}+a; \xi_a)
\nonumber
\ee
\be
dv_{\mathbb{CP}^N}=\frac{1}{(1+y^2)^{N+1}}\prod_{a=1}^N  y_a d y_a d\varphi_a\,,
\label{PvC}\ee
where
\be
\xi_a=\frac{y_a^2}{y_a^2+y_{a+1}^2}.
\ee

\subsection*{Reduction to quantum (spherical) Rosochatius system}
From the above consideration it is clear that, by fixing the eigenvalues of $\widehat{J}_{a\bar a}=\widehat{p}_{\varphi_{a}}$, we can reduce the Hamiltonians \eqref{qH} and \eqref{rqH}
to those of the quantum (spherical) Rosochatius system, the classical counterpart of which is defined by eq. \eqref{realHam}.

However, the quantization  of \eqref{realHam} through replacing the kinetic term by the Laplacian yields a slightly different expression for the Hamiltonian
\be
\widehat{H}_{Ros}=-\frac{\hbar^2}{4}(1+\sum_{c=1}^Ny^2_c)\Bigg[\sum_{a,b=1}^{N}(\delta_{ab}+y_a y_b)\frac{\partial^2}{\partial y_a\partial y_b}+\sum_{a=1}^N\left(2y_a\partial_{y_a}+\frac{g_a^2}{y_a^2}\right)+g_0^2\Bigg].
\label{qsRH}\ee
This is because the volume element on $N$-dimensional sphere is different from that reduced from  $\mathbb{CP}^N$:
\be
dv_{S^N}=\frac{1}{(1+\sum_{c=1}^Ny^2_c)^{(N+1)/2}}\prod_{a=1}^N d y_a ,
\ee
and it gives rise to a different Laplacian as compared to that directly obtained by reduction of the Laplacian on $\mathbb{CP}^N\,$.

As a result,  the relation between wavefunctions of the (spherical) Rosochatius system and those of  $\mathbb{CP}^N$-Rosochatius system is as follows,
\be
\Psi_{sph}=\sqrt{\frac{(1+\sum_{c=1}^Ny^2_c)^{(N+1)}}{{\prod_{a=1}^N y_a}}}\Psi\,.
\ee
So in order to transform the reduced $\mathbb{CP}^N$-Rosochatius Hamiltonian to the spherical one \eqref{qsRH}, we have to redefine the wavefunctions presented in \eqref{PvC}
 and perform the respective similarity transformation of the Hamiltonian.

\section{Supersymmetry}
Let us briefly discuss the possibility of supersymmetrization of $\mathbb{CP}^N$-Rosochatius system, postponing the detailed analysis for a separate study
\cite{inss}.
The $\mathbb{CP}^N$-Rosochatius system belongs to the class of the so-called  ``K\"ahler oscillators" \cite{CPosc,Kahlerosc} (up to a constant shift of the Hamiltonian),
and therefore, admits $SU(2|1)$ (or, equivalently, ``weak $\mathcal{N}=4$") supersymmetric extension.
Namely, its Hamiltonian \eqref{Ham} can be cast in the form
\be
\mathcal{H}_{Ros}=\sum_{a,b=1}^N g^{\bar a b}\left(\bar\pi_a\pi_b + |{\omega}|^2\partial_{\bar a}K\partial_{b}K\right)- |\sum_{i=0}^N\omega_i|^2-\sum_{i=0}^N|\omega_i|^2,
\label{Kosc}\ee
with
\be
 K=\log(1+z\bar z)-\frac{1}{|\omega|}\sum_{a=1}^N ({\omega_a}\log z^a +{{\bar \omega}_a}\log {\bar z}^a),\quad \omega={\omega_0+\sum_{a=1}^N\omega_a}.
\label{suprel}\ee
Here, as opposed to the previous Sections, we assume that $\omega_i$ are complex numbers, {\it i.e.} we replaced
\be
\omega_i\to\omega_i{\rm e}^{\imath\nu_i},
\ee
 with $\nu_i$ being arbitrary real constants.

The $SU(2|1)$ superextension just mentioned is reduced to that with $\mathcal{N}=4, d=1 $  Poincar\'e supersymmetry
under the conditions\footnote{From the viewpoint of $SU(2|1)$ mechanics, $B$ is just the parameter of contraction to $\mathcal{N}=4, d=1$ supersymmetry \cite{sidorov}.}
\be
B=0\,,\quad \omega=\sum_{i=0}^N\omega_i=0.
\label{Bo0}\ee
One could expect that the second constraint corresponds to the vanishing   potential.
 However, it is not the case: looking at the explicit expression for  the Hamiltonian, one can see that the  parameter $\omega$ does not appear
 in denominators anymore. { Indeed, the second constraint above leads to the relation $|\omega_0|^2=|\sum_{a=1}^N\omega_a|^2$, which allows
  to represent  the Hamiltonian \eqref{Ham} in the following form
 \be
 \mathcal{H}_{Ros}=\sum_{a,b=1}^N g^{\bar a b}\left(\bar\pi_a\pi_b +\partial_{\bar a}{\bar U}\partial_{b}U\right)-\sum_{i=0}^N|\omega_i|^2\,,
 \label{hHam}\ee
where  $U(z)$ is  the holomorphic function (``superpotential")
\be
{U}(z)=\sum_{a=1}^N\omega_a\log z^a.
\ee
It is well known that the systems with such a Hamiltonian admit  the $\mathcal{N}=4$ supersymmetric extension  in the absence  of magnetic field (see, e.g., \cite{PRDrapid}).
Explicitly, it amounts to the following consideration.

Let us consider  a $(2N.4N)_{C}$-dimensional  phase space
equipped with the symplectic structure (till the end of this section we assume the summation over repeating indices)
\begin{equation}
\begin{array}{c}
\Omega=d\pi_a\wedge dz^a+ d{\bar\pi}_a\wedge d{\bar z}^a
-\frac{1}{2}R_{a{\bar b}c\bar d}\eta^c_\alpha\bar\eta^{d\alpha}
dz^a\wedge d{\bar z}^b+
\frac12 g_{a\bar b}D\eta^a_\alpha\wedge{D{\bar\eta}^b_\alpha}\quad,
\end{array}
\label{ss}\end{equation}
where $D\eta^a_\alpha
=d\eta^a_\alpha+\Gamma^a_{bc}\eta^b_\alpha dz^c$ with
 $\Gamma^a_{bc},\; R_{a\bar b c\bar d}$ being
respectively, the components of connection and curvature of
the K\"ahler structure associated with the Fubini-Study metrics \eqref{FS}, $\eta^{a}_{\alpha}, \bar\eta^{a}_{\alpha}$ are Grassmann variables with additional
$SU(2)$ indices $\alpha=1,2$. The lower- and upper-case $SU(2)$ indices are related by the antisymmetric matrix $\epsilon_{\alpha \beta}$ and its inverse $\epsilon^{\alpha\beta}$
($\epsilon_{12} = \epsilon^{21}=1$).

The Poisson brackets  defined by \eqref{ss} are given
%\be
%\{f,g\}=\frac{\partial f}{\partial\pi_a}\wedge\nabla_a g +\frac{\partial f}{\partial\bar\pi_a} \wedge{\bar\nabla}_a g +\imath(Bg_{a\bar b}+
%\imath R_{a\bar b c\bar d}\eta^c_\alpha{\bar\eta}^d_\alpha)\frac{\partial f}{\partial\pi_a}\wedge\frac{\partial g}{\partial\bar\pi_b}+g^{a\bar b}\frac{\partial^r f}{\partial\eta^a_\alpha}\wedge\frac{\partial^l g}{\partial\bar\eta^b_\alpha},
%%\ee
%\quad {\rm with}\quad
%%
%\nabla_a\equiv \frac{\partial}{\partial z^a}-\Gamma^c_{ab}\eta^b_\alpha\frac{\partial^r}{\partial \eta^c_\alpha},
%\ee
%and $f\wedge g= fg-(-1)^{p(f)p(g)}gf $.
by the following non-zero
relations and their complex conjugates:
\be
\{\pi_a, z^b\}=\delta^b_a,\quad
\{\pi_a,\eta^b_\alpha\}=-\Gamma^b_{ac}\eta^c_\alpha,\quad
\{\pi_a,\bar\pi_b\}=- R_{a\bar b c\bar d}\eta^c_\alpha{\bar\eta}^{d\alpha},
\quad
\{\eta^a_\alpha, \bar\eta^{b\beta}\}=
g^{a\bar b}\delta_{\alpha}^{\beta}.
\ee
%The symplectic structure \eqref{ss} and respective Poisson brackets
%are manifestly invariant with respect to transformations
%\be
%{\widetilde z}^a={\widetilde z}^a(z), \qquad {\widetilde\pi}_a=\frac{\partial z^b}{\partial{\widetilde z}^a}\pi_b, \qquad{\widetilde\eta}^a_\alpha=\frac{\partial{\widetilde z}^a}{\partial z^b}\eta^b_\alpha.
%\ee
Straightforward calculations show that the following supercharges and Hamiltonian obey the $\mathcal{N}=4, d=1$ Poincar\'e superalgebra
\bea
&Q^\alpha=  \pi_a \eta^{a\alpha}+ \imath {\bar U}_{,\bar a}{\bar \eta}^{a\alpha},\quad {\overline Q}_{\alpha}=
\bar\pi_a \bar\eta^a_\alpha +\imath U_{,a} {\eta}^a_\alpha, &\\
& \mathcal{H}_{SUSY}=\mathcal{H}_{Ros}
-\frac12 R_{a\bar b c\bar d}\eta^{a\alpha}\bar\eta^b_\alpha\eta^{c\beta}\bar\eta^d_\beta
+\frac{\imath}{2} U_{,a;b}\eta^{a\alpha}\eta^b_\alpha+
\frac{\imath}{2} {\bar U}_{,\bar a;\bar b}\bar\eta^{a\alpha}\bar\eta^b_\alpha\quad : &\label{hosup}\\
&\{Q^{\alpha},{\overline Q}_\beta\}=\delta^\alpha_{\beta}\left(\mathcal{H}_{SUSY}+ \sum_{i=0}^N|\omega_i|^2\right),
\qquad\{Q^{\alpha},Q^\beta\}= \{{\overline Q}_{\alpha},{\overline Q}_\beta\}=\{Q^{\alpha},\mathcal{H}_{SUSY}\}=\{ {\overline Q}_{\alpha},\mathcal{H}_{SUSY}\}=0,
\eea
Hence, when the  constraints \eqref{Bo0} are imposed, we can construct $\mathcal{N}=4$ supersymmetric extension of $\mathbb{CP}^N$-Rosochatius system.

An interesting issue is the symmetries of  the supersymmetric system constructed. Writing down the explicit expressions for the Hamiltonian and supercharges one can  be convinced
that they are explicitly invariant under $U(1)$-transformations $z^a\to{\rm e}^{\imath\kappa}z^a, \pi_a\to{\rm e}^{-\imath\kappa}\pi_a, \eta^{a\alpha}
\to{\rm e}^{\imath\kappa}\eta^{a\alpha}$ which are obviously canonical transformations. Hence, one can easily construct the ``supersymmetric counterpart" of $U(1)$  generators
\eqref{u1n}. However, it is still unclear whether hidden symmetries of the  system one started with can be lifted to its supersymmetric extension.
A more detailed analysis of these questions will be a subject of \cite{inss}.

}

Let us emphasize that the restriction $\omega=0$ can be graphically represented  as a planar  polygon  with the edges $|\omega_i|$ (see Fig.1),
which leads to the inequality
\be
|\omega_i|\leq \sum_{j\neq i}|\omega_j|. \label{InEq}
\ee
\begin{tikzpicture}
 \draw[thick,->] (0,0) -- (0,3);

 \draw[thick,<-] (2,0) -- (2,3);

\draw (-0.5,1.7) node {$\omega_0$};
\draw (1.5,1.7) node {$\omega_1$};

  \draw[thick,->] (6,0.3) -- (7.5,3) ;

    \draw[thick,->]  (7.5,3)--(8,0.3) ;

  \draw[thick,->] (8,0.3) -- (6,0.3) ;

\draw (6.2,1.7) node {$\omega_0$};
\draw (8.2,1.7) node {$\omega_1$};
\draw (7,0) node {$\omega_2$};

 \draw[thick,->] (12.5,0.3) -- (13.5,2);

    \draw[thick,->]  (13.5,2)--(15.5,3);

  \draw[thick,->] (15.5,3) -- (16.5,0.3) ;

 \draw[thick,->] (16.5,0.3) -- (12.5,0.3);

\draw (12.8,1.4) node {$\omega_0$};
\draw (14.5,2.8) node {$\omega_1$};
\draw (16.3,1.7) node {$\omega_2$};
\draw (14.5,0) node {$\omega_4$};
\end{tikzpicture}
\vspace{0.2cm}
\begin{center}
Fig.1
\end{center}
\noindent This implies that:
\begin{itemize}
\item For $N=1$  the constraint $\omega=0$ uniquely fixes the values of parameters in the case of $\mathbb{CP}^1$: $ \nu_0=-\nu_1$  and $|\omega_0|=|\omega_1|$.
 The latter property leads to the appearance of discrete symmetry
\be
z\to \frac{1}{z}.
\label{discrete}
\ee

\item For  $N=2$ the above constraints amount to a triangle, which  fixes the parameters $\nu_a$ as follows
\be
\cos{(\nu_2-\nu_0)}=\frac{|\omega_1|^2-|\omega_0|^2 -|\omega_2|^2}{2|\omega_0||\omega_2|},\quad\cos{(\nu_1-\nu_0)}
=\frac{|\omega_2|^2-|\omega_0|^2 - |\omega_2|^2}{2|\omega_0||\omega_1|}\,.\quad
\ee
\item  For $N>2$ the parameters $\nu_a$ are not uniquely fixed, so that we obtain a family of $\mathcal{N}=4$ supersymmetric Hamiltonians depending on up to $N-1$ parameters.
\end{itemize}
We observe that for any value of $N$  at least one parameter  $\nu_i$ remains unfixed. But this does not affect our consideration  since
such parameter can be absorbed into a redefinition of  fermionic variables.

Finally, note that
the constraint $\sum_{i=1}^N\omega_i=0$  also appeared  in constructing $\mathcal{N}=4$ supersymmetric extension of  $\mathbf{S}^N$-Rosochatius system
\cite{wdvv}, but  with $\omega_i$ being real numbers. The above trick   with complexification of the parameters $\omega_i$ is {seemingly }
applicable to the $\mathbf{S}^N$-Rosochatius system as well, hopefully giving rise to a less restrictive form of the $\mathcal{N}=4$ superextension of the latter.

\section{Concluding remarks}
In this paper we proposed  the superintegrable $\mathbb{CP}^N$-analog of Rosochatius and Smorodinsky-Winternitz systems  which is specified by the presence of constant magnetic field
and is form-invariant under transition from one chart of $\mathbb{CP}^N$ to others accompanied by the appropriate permutation  of the characteristic parameters $\omega_i$.
We  showed that the system  possesses $3N-1$ functionally independent constants of motion
and explicitly constructed its classical and quantum solutions. In the generic case this model admits an extension with $SU(2|1)$ supersymmetry,
which is reduced,  under the special choice
of the characteristic parameters and in the absence of magnetic field, to the ``flat'' $\mathcal{N}=4, d=1$ Poincar\'e' supersymmetry.

 When all constants $\omega_i$ are equal, the system is covariant under the above transitions between charts and so becomes globally defined
on the whole $\mathbb{CP}^N$ manifold. This covariance implies $N$ discrete symmetries,
\be
z^a\to \frac{1}{z^a},\quad z^\alpha\to\frac{z^\alpha}{z^a},\quad {\rm with }\quad\alpha\neq a.
\ee
Moreover, in this special case the model always admits (in the absence of magnetic field) $\mathcal{N}=4, d=1$ Poincar\'e' supersymmetrization because the inequality \eqref{InEq}
is automatically satisfied.
The model with equal $\omega_i$ can be also interpreted as  a model of $N$ {\sl interacting } particles with an effective position-dependent mass
located in the quantum ring.  This agrees with the property that, in the flat limit, the model under consideration  can be interpreted  as an ensemble of  $N$  free
particles  in a single quantum ring interacting with a constant magnetic field orthogonal to the plane (cf. \cite{chp,chp1,chp2}).
Thus the property of the exact solvability/superintegrability of the suggested model in the presence of constant magnetic field (equally as of the superextended
model implying the appropriate
inclusion of spin) makes it interesting  also from this point of view.

The obvious next tasks are the study of {classical and quantum $SU(2|1)$  supersymmetric extension of the  $\mathbb{CP}^N$ Rosochatius system \cite{inss}, as well
as the  construction of its Lax pair formulation.}

Two important possible generalizations of the proposed system are the following ones:
\begin{itemize}
\item  An analog of $\mathbb{CP}^N$-Rosochatius system  on the quaternionic projective space $\mathbb{HP}^N$ in the presence of BPST instanton.

    Presumably, it can be defined by the Hamiltonian \eqref{Hamr0} and the symplectic structure \eqref{hn}, in which  $\pi_a,  z^a$ are replaced by quaternionic variables,
    and  the last term  in \eqref{hn} by terms responsible  for interaction
    with BPST instanton \cite{duval} (see also \cite{KoSmi}, \cite{IKoSmi} and \cite{QPN}). The phase space of this system is expected to be
    ${T}^*\mathbb{HP}^N\times \mathbb{CP}^1$, due to the isospin nature of instanton. We can hope that
    this system is also superintegrable and that  an interaction with BPST instanton  preserves the superintegrability. On this way we can also expect
    intriguing links with the recently explored  Quaternion-K\"ahler deformations of ${\cal N}=4$ mechanics \cite{ILuc}. These models also admit homogeneous
    $\mathbb{HP}^N$ backgrounds.

\item   $\mathbb{CP}^N$-analog of Coulomb problem.

     Such an extension could  be possible, keeping in mind the existence of superintegrable spherical analog  of Coulomb problem with
   additional $\sum_i g^2_i/x^2_i$  potential, as well as the observation that  the (spherical) Rosochatius system  is a real section of
 $\mathbb{CP}^N$-Rosochatius system.

\end{itemize}
One of the key  motivations of the present study  was to derive the superintegrable $\mathbb{CP}^N$- and $\mathbb{C}^N$- generalizations
of rational Calogero model. Unfortunately, until now we succeeded in constructing only trivial extensions of such kind.
We still hope to reach the general goal just mentioned in the future.

\acknowledgments
We thank Levon Mardoyan, George Pogosyan and Stepan Sidorov for useful comments and Tigran Hakobyan and  Sergey Vinitsky for encouraging.
E.I. acknowledges
 a partial support from the RFBR grant, project No 18-02-01046.
 The work of A.N. and H.S. was fulfilled within the ICTP Affiliated Center Program AF-04  and ICTP Network project NT-04
. They both acknowledge financial support from the grants 18RF-002 and 18T-1C106 of  Armenian Committee of Science. H.S.  acknowledges
 FAST-Foundation for Armenian Science and Technology, the Regional Doctoral Program on Theoretical and Experimental Particle Physics Program
 sponsored by VolkswagenStiftung  and ANSEF-Armenian National Science and Education Fund based in New York  for financial support.

{ We thank the anonymous Referee for carefully reading of the manuscript and pointing out many misprints, as well as for drawing
our attention to the highly interesting series of papers by M.Plyushchay with co-authors  on non-linear supersymmetric extensions
of finite-gap systems (see, e.g., \cite{Misha} and refs. therein). For sure, this topic, in the context of our present study,
deserves a separate consideration}.

  \end{document}